\documentclass[12pt]{JHEP3}
\usepackage{amsmath,amssymb,bbm}
\def\de{\partial}

\def\2{\frac12}
\def\4{\frac14}
\def\ie{{\it i.e.}~}
\def\eg{{\it e.g.}~}

\newcommand{\be}{\begin{equation}}
\newcommand{\ee}{\end{equation}}
\newcommand{\bea}{\begin{eqnarray}}
\newcommand{\eea}{\end{eqnarray}}
\newcommand{\ba}{\begin{array}}
\newcommand{\ea}{\end{array}}
\def\a{\alpha}
\def\b{\beta}

\def\d{\delta}
\def\e{\epsilon}

\def\k{\kappa}
\def\l{\lambda}
\def\L{\Lambda}
\def\m{\mu}
\def\n{\nu}

\def\r{\rho}
\def\s{\sigma}

\def\de{\partial}

\def\ad{{\dot \alpha}}
\def\bd{{\dot \beta}}
\def\cA{{\cal A}}

\author{Massimo Bianchi\\Dipartimento di Fisica, Universit{\`a} di Roma ``Tor
Vergata'', I.N.F.N. - Sezione di Roma II ``Tor Vergata'',Via della
Ricerca Scientifica, 1 - 00133 Roma - ITALY\\
\email{Massimo.Bianchi@roma2.infn.it}}
\author{Paul J. Heslop and Fabio Riccioni\\DAMTP,
Centre for Mathematical Sciences, University of Cambridge,
Wilberforce Road, Cambridge CB3 0WA,  UK\\ \email{P.J.Heslop,
F.Riccioni@damtp.cam.ac.uk}} 
\abstract{We discuss higher spin gauge
symmetry breaking in AdS space from a holographic prespective.
Indeed, the AdS/CFT correspondence implies that ${\cal N}=4$ SYM
theory at vanishing coupling constant is dual to a theory in AdS
which exhibits higher spin gauge symmetry enhancement. When the SYM
coupling is non-zero, the current conservation condition becomes
anomalous, and correspondingly the local higher spin symmetry in the
bulk gets spontaneously broken. In agreement with previous results
and holographic expectations, we find that the Goldstone mode
responsible for the symmetry breaking in AdS has a non-vanishing
mass even in the limit in which the gauge symmetry is restored.
Moreover, we show that the mass of the Goldstone mode is exactly the
one predicted by the correspondence. Finally, we obtain the precise
form of the
higher spin supercurrents in the SYM side.} \preprint{ROM2F-05/08 \\
DAMTP-2005-40 \\ hep-th/0504156}

\keywords{AdS-CFT and dS-CFT Correspondence,  Supersymmetry and
  Duality,  Space-Time Symmetries}

\title{More on {\it La  Grande  Bouffe:}
  towards \\ higher spin symmetry breaking in AdS}

\begin{document}

\section{Introduction and summary}

The AdS/CFT correspondence \cite{adscft,adscftrev} between ${\cal N}
=4$ SYM and type IIB superstring on $AdS_5 \times S^5$ has received
a new wave of attention in the past few years. On the one hand, the
BMN limit has suggested the possibility that some `unprotected'
observables on the two sides of the correspondence could be matched
in particular regimes \cite{bmnetal,gkpspin}. On the other hand,
exploiting Higher Spin (HS) symmetry enhancement at small radius
\cite{smallradius, hsinads4} and taking the BMN limit as a hint, it
is possible to extend the results of
\cite{sezsund,sezsund2,konvaszai} for the `first Regge trajectory'
to the full string spectrum, showing that it precisely matches with
the operator spectrum of free ${\cal N} =4$ SYM
 in the planar limit~\cite{bbms1,bbms2}. When interactions are turned
on, \ie at finite radius but still at vanishing string coupling, all
but a handful of higher spin string states become massive by eating
lower spin `Goldstone' fields. This pantagruelic Higgs mechanism,
termed {\it La Grande Bouffe} in \cite{bms}, is the bulk counterpart
of the anomalous violation of classically conserved currents in the
boundary CFT. The resolution of the operator mixing yielding the
anomalous dimensions \cite{opemix} has been shown to be equivalent
to diagonalizing a dilatation operator \cite{dilatop} that in turn
can be identified with the Hamiltonian of a supersymmetric spin
chain \cite{spinchain}. Integrability of the system at higher loops
turns out to be much subtler then expected since it naively leads to
inconsistencies with holography \cite{integrable}. Yet the
encouraging results at the Higher Spin enhancement point suggest
that Higher Spin Symmetry
\cite{sezsund,sezsund2,konvaszai,vasiliev,hs,brux} could shed some
light on this issue and at least help organize {\it La Grande
Bouffe}. In particular decomposing Higher Spin multiplets into
superconformal multiplets allows one to combine the latter into long
multiplets as expected for unprotected massive states
\cite{bbms1,bbms2}.

The aim of this paper is to give a detailed account of the bulk
counterpart of this mechanism for totally symmetric spin $s$
fields at the quadratic level in the lagrangian. Although this may
sound kinematical, and it is indeed so in so far as mass shifts
are left undetermined, we believe Higher Spin Symmetry should
constrain if not completely fix the mass shifts. At least it
should relate shifts for different spins in the same HS multiplet,
very much as group theory fixes the ratios of the masses of the
vector bosons after spontaneous breaking of a gauge symmetry. HS
currents $J_{i_1 \dots i_s}$ with $s >2$ occur in ${\cal N}=4$ SYM
where they are conserved and belong to semishort multiplets at
vanishing coupling $g=0$. Interactions are responsible for their
anomalous violation
$$ \de^{i_1} J_{i_1 \dots i_s} = g {\cal X}_{i_2 \dots
i_s} \quad ,
$$
where the anomalous term ${\cal X}_{i_2\dots i_s}$ lies in a
different supermultiplet with respect to $J_{i_1 \dots i_s}$ at
vanishing coupling. Conformal invariance fixes the dimension of such
a spin $s$ conserved current on the $d$ dimensional boundary to be
$s+d-2$, when the coupling vanishes. In the same limit the dimension
of the anomalous term is $s+d-1$. This implies that ${\cal X}$ is
not a conserved spin $s-1$ current when $g=0$, and therefore one
expects it to be dual to a massive field in the bulk. With this in
mind, we therefore study the Higgs mechanism for totally symmetric
spin $s$ fields in AdS. Our results are in agreement with the
results of~\cite{DESWALD}, where `partial masslessness' in (A)dS was
studied in great detail\footnote{We thank S. Deser and A. Waldron
for pointing out to us their and Zinoviev's~\cite{ZINO}
contributions to the subject.}. We will first describe the Higgs
mechanism in flat spacetime, as discussed in~\cite{tesifabio} and
briefly reviewed in~\cite{mbcreta, mbstring}. We adopt the
St\"uckelberg description, whereby a massless spin $s$ field eats a
massless spin $s-1$ field, which in turn eats a massless spin $s-2$
field and so on up to $0$ (or $1/2$ if the initial field is a
fermion). In flat spacetime, when the gauge symmetry is restored,
{\it all} the Goldstone modes become massless, {\it i.e.} all the
fields from spin $s$ to spin $1$ become gauge fields. Holography
implies that the same Higgs mechanism works differently in AdS,
where the `Goldstone' spin $s-1$ field becomes a massive field when
the spin $s$ gauge symmetry is restored \cite{ZINO}.

The paper is organized as follows. In section 2 we consider spin 1
fields, for which the flat space case and the AdS case actually
coincide. We then move in section 3 to spin 2, which for our
(holographic) purposes is not to be thought of as the graviton but
rather as one of the two extra symmetric tensors that appear in the
superstring spectrum at the HS enhancement point (one belongs to the
Konishi multiplet and is a superdescendant at level four, while the
other is the primary of a spin 2 semishort multiplet). The Higgs
mechanism for the graviton in $AdS_4$ was discussed by Porrati in a
very interesting and suggestive series of papers \cite{porrati}. The
crucial point is that in AdS a massive spin 2 field decomposes into
a massless spin 2 and a {\it massive} spin 1~\cite{DESWALD,ZINO}. We
then extend the analysis to the case of arbitrary symmetric tensors
in section 4, showing in particular that in AdS a massive spin $s$
field decomposes into a massless spin $s$ and a massive spin $s-1$.
Moreover, we show that the mass of the spin $s-1$ Goldstone field is
exactly in agreement with the one predicted by the AdS/CFT
correspondence. Unfortunately the most general case in AdS for $d\ge
4$ are mixed symmetry tensors that can become `massless'
\cite{metsaevmixsym, metsmass} when they are part of (semi)short
multiplets. In order to cope with these, in section 5 we reformulate
our kinematical analysis in manifestly supersymmetric terms by
carefully identifying the boundary operators and their dual bulk
fields participating in the shortening of superconformal multiplets
at threshold, \ie when interactions are turned off and the AdS
radius is `small' and comparable with the string scale. Finally we
conclude with remarks and perspectives for future investigation.

\section{Spin one: Higgs \'a la St\"uckelberg}

A very convenient way of describing a massive vector field is
through the St\"uckelberg formulation. One starts with a massless gauge
fields $A_\mu$ and a massless scalar field $\chi$ with an `axionic'
Peccei-Quinn symmetry $\chi \rightarrow \chi + \alpha_0$, where
$\alpha_0$ is an arbitrary constant and then `gauges' the symmetry
by means of $A_\mu$. In formulae
\be \delta A_\mu = \de_\mu \alpha
\quad , \qquad \delta\chi = m \alpha \quad .\label{gaugeinvone} \ee
In flat
spacetime of any dimension $D$, a Lagrangian invariant under
(\ref{gaugeinvone}) is
\be {\cal L} = -{1\over
4} F_{\mu\nu}F^{\mu\nu} - {1\over 2} (\de_\mu \chi - m
A_\mu)(\de^\mu \chi - m A^\mu) \quad ,\label{spinoneflat} \ee where
$F_{\mu\nu}= \de_\mu A_\nu - \de_\nu A_\mu$ is gauge invariant by
construction. Clearly for $m=0$, $\chi$ and $A_\mu$ simply decouple
and are both massless, while for $m\neq 0$ one can `gauge $\chi$
away' by means of (\ref{gaugeinvone}). As a result the gauge
invariant vector field ${\cal A}_\mu = A_\mu - {1\over m} \de_\mu
\chi$ is a massive spin one field with the correct number of degrees
of freedom $\nu_{s=1}^{m\neq 0}  = D-1 = \nu_{s=1}^{m = 0} +
\nu_{s=0}^{m=0} = D-2 + 1$.

The nice feature of the above construction is that it admits a straightforward
generalization to any curved background without torsion. One simply has to
replace partial ($\de_\mu$) with covariant ($\nabla_\mu$) derivatives and to
contract indices with the relevant background metric tensor $g_{\mu\nu}$.
The crucial point is that the definition of
$F_{\mu\nu}$ is unchanged with respect to its flat spacetime expression.

The Lagrangian
\be
{\cal L} = -{1\over 4} F_{\mu\nu}F^{\mu\nu} -
{1\over 2} (\nabla_\mu \chi - m A_\mu)(\nabla^\mu \chi - m A^\mu)
\label{spinonecurved}
\ee
is manifestly gauge invariant under
\be
\delta A_\mu = \nabla_\mu \alpha \equiv \de_\mu \alpha \quad , \qquad
\delta\chi = m \alpha \quad .
\label{gaugeinvonecurved}
\ee

To conclude, let us put our findings in a holographic perspective.
A massless vector field $A_\mu$ in the AdS$_{d+1}$ bulk
corresponds to a dimension $\Delta_1 = d - 1 = 1 + d-2$
classically conserved current $J_i$ in the boundary CFT$_d$, that
lives at the unitary bound for spin $s=1$ irreps of $SO(d-2,2)$. A
massless scalar field $\chi$ in the AdS$_{d+1}$ bulk corresponds
to a dimension $\Delta_0 = d = \Delta_1 + 1$ marginal operator
${\cal X}$ in the boundary CFT$_{d}$. The anomalous violation of
the current $\de^i J_i = g {\cal X}$ is the boundary counterpart
of the Higgs mechanism in the bulk. Two comments are in order.
First one can violate a current by a relevant deformation (mass
term) of the fixed point boundary action. This is indeed what
happens in many interesting RG flows that admit a holographic
description in terms of (supersymmetric) domain wall solutions in
the bulk \cite{holoren}. These solutions are only asymptotic to
AdS, meaning that (super)conformal invariance is broken
explicitly, by deforming the action, or spontaneously, by turning
on the VEV's of any (scalar) operator. Here we are interested in
HS symmetry breaking patterns compatible with (super)conformal
invariance. This is possible only if the free boundary CFT is
deformed by an exactly marginal deformation. In ${\cal N}=4$ this
corresponds to turning on the (complexified) gauge coupling. All
other potentially marginal (single trace\footnote{Exactly marginal
double-trace deformations of the Leigh-Strassler type
\cite{leighstras} are known to exist \cite{d20} and their effects
have been studied in \cite{aharonberk}.}) deformations are not
integrable, \ie they acquire anomalous dimensions when they are
used to perturb the theory away from the free field theory limit.
Second, the masslessness of the St\"uckelberg field $\chi$ is an
accident of the spin one case. For spin higher than one in AdS,
holography or, more simply, $SO(d-2,2)$ group theory suggest that
the relevant spin $s'=s-1$ St\"uckelberg field be massive. This is
precisely what we are going to see in detail in the next two
sections.

\section{Spin two}

\subsection{$s$=2 in flat spacetime}

The lagrangian describing the St\"uckelberg form of a massive spin
2 free field in flat space-time in $D$ dimensions can be formally
obtained starting from a massless spin 2 field in $D+1$
dimensions, described by the lagrangian \be {\cal L} =-\frac{1}{2}
(\de_M \Phi_{NR})^2 + (\de_N \Phi^N{}_M )^2 +\frac{1}{2} (\de_M
\Phi^L{}_L)^2 + \Phi^L{}_L \de_M \de_N \Phi^{MN}  \quad , \ee and
allowing the field to depend harmonically on the $D+1$-th
coordinate, \be \Phi (x,y) = \frac{1}{\sqrt2}\Phi(x) e^{imy} +
{\rm c.c.} \quad . \ee After redefining the $D$-dimensional
fields, \be \Phi_{\m\n} = \phi_{\m\n} \quad , \quad \Phi_{\m y} =
-i \chi_\m \quad , \quad \Phi_{yy} = \psi \quad , \ee where
$\phi_{\m\n}$, $\chi_\m$ and $\psi$ are chosen to be real, one
obtains the lagrangian \bea {\cal L} =& & -\frac{1}{2} (\de_\m
\phi_{\n\r})^2 + (\de_\n \phi^\n{}_\m )^2 +\frac{1}{2} (\de_\m
\phi^\l{}_\l)^2 + \phi^\l{}_\l
\de_\m \de_\n \phi^{\m\n} \nonumber \\
& & - \frac{m^2}{2} \phi_{\m\n}^2 +\frac{m^2}{2}( \phi^\l{}_\l )^2 -
(\de_\m \chi_\n )^2 + (\de_\m \chi^\m)^2 \nonumber \\
& & +\de_\m \psi \de^\m \phi^\l{}_\l +\psi \de_\m \de_\n \phi^{\m\n}
+ 2m \chi_\m \de_\n \phi^{\m\n} +2 m \phi^\l{}_\l \de_\m \chi^\m \
\quad , \label{lagrangian} \eea that is invariant with respect to
the gauge transformations \be \d \phi_{\m\n} = 2 \de_{(\m} \e_{\n)}
\quad , \quad \d \chi_\m = \de_\m \alpha - m \e_\m \quad , \quad \d
\psi = 2 m \alpha \quad . \label{gaugetransfflat}\ee If $m$ is
non-vanishing, one can use both $\e_\m$ and $\alpha$ to gauge away
$\chi_\m$ and $\psi$, so that the resulting lagrangian describes a
massive spin 2 field. If $m$ vanishes instead, the lagrangian
describes massless spin 2, spin 1 and spin 0 fields\footnote{A
redefinition of $\phi_{\m\n}$ of the form $\phi_{\m\n} =
\phi_{\m\n}^\prime - \frac{1}{D-2} \eta_{\m\n} \psi$ is needed in
order to obtain the standard kinetic term for the scalar, so that in
the lagrangian (\ref{lagrangian}) the fields decouple for $m=0$.}.

We will see in the next subsection which changes are needed when one
considers AdS instead of flat spacetime. In this direction, it is
very instructive to prove explicitly gauge invariance of the
lagrangian (\ref{lagrangian}) with respect to the transformations
(\ref{gaugetransfflat}), and it is actually easier to consider the
field equations \bea & & \Box \phi_{\m\n} -2 \de_{(\m} (\de\cdot
\phi)_{\n)} + \de_\m \de_\n \phi^\l{}_\l
- \eta_{\m\n} [ \Box \phi^\l{}_\l - \de \cdot \de \cdot \phi ]
\nonumber \\
& & \ - m^2 \phi_{\m\n} + m^2 \eta_{\m\n}\phi^\l{}_\l -2m \de_{(\m}
\chi_{\n)} + 2 m \eta_{\m\n} (\de \cdot \chi ) -
\eta_{\m\n} \Box \psi + \de_\m \de_\n \psi =0 \ , \label{2flat} \\
&& \Box \chi_\m - \de_\m (\de \cdot \chi ) + m (\de\cdot \phi)_\m - m
\de_\m \phi^\l{}_\l =0 \quad , \\
&& \Box \phi^\l{}_\l - \de_\m \de_\n \phi^{\m\n} =0 \quad ,
\label{0flat} \eea obtained varying (\ref{lagrangian}) with respect
to $\phi_{\m\n}$, $ \chi_\m$ and $\psi$ respectively, instead of the
lagrangian itself.

If $m \neq 0$, one is allowed to perform the redefinition of
$\chi_\m$\be \chi^\prime_\m = \chi_\m - \frac{1}{2m} \de_\m \psi
\label{redef}\quad , \ee such that the gauge transformation of the
new field does not contain $\a$ anymore: \be \d \chi^\prime_\m = - m
\e_\m \quad . \ee In terms of the new field (we will drop the
`prime' index from now on), the equations are \bea && \Box
\phi_{\m\n} -2 \de_{(\m} (\de\cdot
\phi)_{\n)} + \de_\m \de_\n \phi^\l{}_\l
- \eta_{\m\n} [ \Box \phi^\l{}_\l - \de \cdot \de \cdot \phi ]
\nonumber \\
&& \quad - m^2 \phi_{\m\n} + m^2 \phi^\l{}_\l -2m \de_{(\m} \chi_{\n)} +
2 m \eta_{\m\n} (\de \cdot \chi )=0 \quad , \label{spin2flat} \\
&& \Box \chi_\m -\de_\m (\de \cdot \chi ) + m (\de\cdot \phi)_\m - m \de_\m
\phi^\l{}_\l =0 \quad , \label{spin1flat} \\
&& \Box \phi^\l{}_\l - \de_\m \de_\n \phi^{\m\n} =0 \quad
.\label{spin0flat} \eea There is no field varying with respect to
$\a$, while the gauge invariance with respect to $\e_\m$ is
straightforward. This field redefinition is nothing but a gauge
transformation that removes $\psi$. An analogous gauge
transformation can be used to remove $\chi_\m$, so that once all
gauge invariances are fixed one is left with a massive spin 2 field.
The reason why we perform explicitly this field redefinition for
$\chi_\m$ only (or for the field of rank $s-1$ only when we consider
the spin $s$ case in the next section) is related to our expectation
that $\chi_\m$ become massive in the AdS case, as we are going to
show in the next subsection.

If $m=0$, the gauge parameters can not be used to gauge away $\psi$
and $\chi_\m$ anymore. Correspondingly, there is an inconsistency in
the redefinition of $\chi_\m$ in eq. (\ref{redef}), which appears in
the form of gauge symmetry enhancement, \be \d \chi_\m = \de_\m
\a^\prime \ee in eq. (\ref{spin1flat}). As we will see, this is the
main difference with respect to the AdS case: if the curvature is
non-vanishing, a $1/L$ mass term for $\chi_\m$ appears\footnote{This
is a proper mass term, in the sense that it breaks gauge invariance.
Recall that in AdS gauge invariant field equations require a
mass-like term, which we refer to as the AdS mass in the
following.}, so that the field redefinition removing the gauge
variation with respect to $\a$ continues to be consistent at $m=0$,
since there is no gauge symmetry enhancement.

\subsection{$s$=2 in AdS}

Now we move on to the AdS case. We want to modify the equations
(\ref{spin2flat}), (\ref{spin1flat}) and (\ref{spin0flat}) in such a
way that they will turn out to be invariant with respect to the
gauge transformations \be \d \phi_{\m\n} = 2 \nabla_{(\m}
\e_{\n)} \quad , \quad \d \phi_\m  =-m \e_\m \quad . \ee
The equations will be modified because the commutator of two
covariant derivatives in AdS is \be [ \nabla_\m , \nabla_\n ] V_\r =
- R^\l{}_{\r\m\n} V_\l = \frac{1}{L^2} ( g_{\n\r} V_\m - g_{\m\r}
V_\n ) \quad . \ee

Let us consider eqs. (\ref{spin2flat}) and (\ref{spin0flat}) first.
The variation of the third line of eq. (\ref{spin2flat}) does not
involve any commutator, while the first two lines, as well as the
whole of eq. (\ref{spin0flat}), become gauge invariant adding a
$1/L^2$ AdS mass term for $\phi_{\m\n}$. The result is \bea && \Box
\phi_{\m\n} -2 \nabla_{(\m} (\nabla \cdot
\phi)_{\n)} + \nabla_{(\m} \nabla_{\n)} \phi^\l{}_\l +
\frac{2}{L^2}(\phi_{\m\n} - g_{\m\n} \phi^\l{}_\l )
 - m^2 \phi_{\m\n} + m^2  g_{\m\n} \phi^\l{}_\l  \nonumber \\
&& \ -2m \nabla_{(\m} \chi_{\n)} + 2 m g_{\m\n} (\nabla \cdot \chi )
- g_{\m\n} \left[ \Box \phi^\l{}_\l - \nabla \cdot \nabla \cdot \phi
- \frac{D-1}{L^2}\phi^\l{}_\l \right]=0 \ , \label{spin2ads} \\
&&\Box \phi^\l{}_\l - \nabla_{(\m} \nabla_{\n)} \phi^{\m\n} -
\frac{D-1}{L^2} \phi^\l{}_\l =0 \quad .\label{spin0ads} \eea Now
consider eq. (\ref{spin1flat}). Observe first of all that if we
write the kinetic term (first two terms) in the form \be \nabla_\n
F^\n{}_\m = \Box \chi_\m - \nabla_\n \nabla_\m \chi^\n \quad , \ee
where we have commuted the indices $\m$ and $\n$ in the derivatives,
this term is gauge invariant with respect to $\d \chi_\m = \nabla_\m
\a$ without the addition of any AdS mass term. In other words, if we
start with this equation and we perform a field redefinition of
$\chi_\m$ like in eq. (\ref{redef}), we are not going to produce
terms containing $\psi$. This means that we just have to consider
the expression
\be \Box \chi_\m -\nabla_\n \nabla_\m \chi^\n + m
(\nabla \cdot \phi)_\m - m \nabla_\m \phi^\l{}_\l \quad ,
\label{spin1ads}\ee
and see whether it is invariant with respect to
the gauge parameter $\e_\m$ or not.

The variation of eq. (\ref{spin1ads}) is \be -2m [\nabla_\m ,
\nabla_\n ] \e^\n = - \frac{2m}{L^2} (D-1) \e_\m \quad , \ee so that
the gauge invariant field equation is \be \Box \chi_\m -\nabla_\n
\nabla_\m \chi^\n + m (\nabla \cdot \phi)_\m - m \nabla_\m
\phi^\l{}_\l - \frac{2(D-1)}{L^2} \chi_\m =0 \quad .
\label{spin1ads2}\ee We therefore obtain a mass term for $\chi_\m$.
Nothing changes if $m \neq 0$, since we can gauge away $\chi_\m$ and
end up with a massive spin 2 field. When $m=0$ instead, the spin 2
field becomes massless, as one can see from eq. (\ref{spin2ads}),
while eq. (\ref{spin1ads2}) still describes a massive spin 1 field
when $m=0$, and there is no gauge symmetry enhancement, unlike the
flat spacetime case. We therefore end up with a massless spin 2 and
with a massive spin 1~\cite{DESWALD,ZINO}.

The same technique can be applied to any spin $s$ (symmetric tensor
with $s$ spacetime indices), to show that in AdS a massive spin $s$
field decomposes into a massless spin $s$ and a massive spin $s-1$
in the limit of vanishing spin $s$ mass. This is what we are going
to show in the next section.

\section{Arbitrary Spin}

\subsection{Any $s$ in flat spacetime}
In order to derive the St\"uckelberg formulation of a massive spin
$s$ field in $D$ dimensions, we proceed in a way similar to the spin
2 case of the previous section. We consider a massless spin $s$
field in $D+1$ dimensions, described in terms of a symmetric rank
$s$ tensor $\Phi_{M_1...M_s}$ ($M_i$ are $D+1$-dimensional spacetime
indices) satisfying the condition \be \Phi^L{}_L{}^M{}_M{}_{...}=0
\label{doubletrace} \quad , \ee and whose lagrangian
\cite{FRONSDAL}\bea {\cal L}= & & - \frac{1}{2}(\de_M
\Phi^{(s)}_{...})^2+\frac{s}{2}(\de\cdot\Phi^{(s)}_{...})^2
+\frac{s(s-1)}{4}(\de_M \Phi^{(s)L}{}_L{}_{...})^2\nonumber\\
& & +
\frac{s(s-1)(s-2)}{8}(\de\cdot\Phi^{(s)L}{}_L{}_{...})^2+\frac{s(s-1)}{2}
\Phi^{(s)L}{}_L{}_{...} (\de\cdot\de\cdot\Phi^{(s)}_{...})
\label{lags}\eea is invariant with respect to the gauge
transformations\footnote{Parentheses $(...)$ denote symmetrization
of spacetime indices with strength one.}
\be
\delta\Phi_{M_1...M_s}=s\de_{(M_1}\e_{M_2...M_s )}\quad ,
\label{delta5} \ee where $\e$ is symmetric traceless with $(s-1)$
indices. As we did in the previous section, we want to obtain the
equations for a massive field in $D$ dimension in the St\"uckelberg
formulation by means of a KK dimensional reduction. We therefore
consider the field to depend harmonically on the $D+1$-th
coordinate, \be \Phi_{\m_1...\m_{s-k}y...y} (x,y)=
(i)^k\phi^{(s-k)}_{\m_1...\m_{s-k}}(x) e^{imy} + {\rm c.c.}\quad,
\label{realfields} \ee and by taking linear combinations we can
choose the fields $\phi^{(s-k)}$ to be real in $D$ dimensions. It is
convenient to perform this KK reduction directly on the equation of
motion, unlike the previous section, where the reduction was
performed on the lagrangian\footnote{This is the reason why the
equations we obtain in this section for $s=2$ are apparently
different to the equations of the previous section. One can check
that they are equivalent for instance by adding eq. (\ref{0flat}) to
the trace of eq. (\ref{2flat}).}.

We thus consider the equation \be \Box\Phi_{M_1...M_s}
-s\de_{(M_1}(\de\cdot\Phi)_{M_2...M_s)}
+\frac{s(s-1)}{2}\de_{(M_1}\de_{M_2}
\Phi^L{}_L{}_{M_3...M_s)}=0\quad , \label{boseq5} \ee describing a
massless spin $s$ field in $D+1$ dimensions, obtained varying the
lagrangian of eq. (\ref{lags}). We want to consider the dimensional
reduction of this equation to $D$ dimensions.  One gets \bea & & [
\Box-\frac{(k-1)(k-2)}{2}m^2]\phi^{(s-k)}_{\m_1...\m_{s-k}}- (s-k)
\de_{(\m_1}(\de\cdot\phi^{(s-k)})_{\m_2...\m_{s-k})} \nonumber \\
& &- (k-1)(s-k)m\de_{(\m_1} \phi^{(s-k-1)}_{\m_2...\m_{s-k})} - k
m(\de\cdot\phi^{(s-k+1)})_{\m_1...\m_{s-k}}\nonumber\\
& & + \frac{(s-k)(s-k-1)}{2}\de_{(\m_1}\de_{\m_2}
\phi^{(s-k)\l}{}_\l{}_{\m_3...\m_{s-k})}-\frac{(s-k)(s-k-1)}{2}\de_{(\m_1}\de_{\m_2}
\phi^{(s-k-2)}_{\m_3...\m_{s-k})}\nonumber\\
\label{ddim} & & +k (s-k) m
\de_{(\m_1}\phi^{(s-k+1)\l}{}_\l{}_{\m_2...\m_{s-k})}+
\frac{k(k-1)}{2}m^2\phi^{(s-k+2)\l}{}_\l{}_{\m_1...\m_{s-k}}=0\quad,
\eea where $k=0,1,...,s$,  and from the $D+1$-dimensional gauge
transformation of eq. (\ref{delta5}) after having defined the $D$
dimensional gauge parameters \be
\e_{\m_1...\m_{s-k-1}y...y}=(i)^k\e^{(s-k-1)}_{\m_1...\m_{s-k-1}}
\label{doubletraceps}\quad, \ee one obtains the gauge
transformations of the $D$-dimensional fields $\phi^{(s-k)}$, \be
\delta\phi^{(s-k)}_{\m_1...\m_{s-k}}=(s-k)\de_{(\m_1}\e^{(s-k-1)}_{\m_2...\m_{s-k})}+
k m\e^{(s-k)}_{\m_1...\m_{s-k}}\quad. \label{fourpointeight} \ee The
traceless condition for $\e$ becomes \be
\e^{(s-k)\l}{}_\l{}_{...}-\e^{(s-k-2)}_{...}=0\quad,
\label{epsilon}\ee while the constraint (\ref{doubletrace}) becomes
\be
\phi^{(s-k)\l}{}_\l{}^\r{}_\r{}_{...}-2\phi^{(s-k-2)\l}{}_\l{}_{...}+
\phi^{(s-k-4)}{}_{...}=0\qquad(k=0,...,s-4)\quad.
\label{fourpointten} \ee

If $m \neq 0$, some of the lower spin fields can be put to zero
fixing their gauge invariance, while some others can not be gauged
away because of the constraint (\ref{epsilon}) on the gauge
parameters. These fields, that are identically zero on shell, are
the auxiliary fields of the massive theory, and one ends up with an
equation for a massive spin $s$ field $\phi^{(s)}$.

The occurrence of auxiliary fields in lagrangians for massive spin
$s$ fields can be understood considering the example of a massive
spin 2 field, whose field equation is \be (\Box -
m^2)\Phi_{\m\n}=0\quad,\qquad(\de\cdot\Phi)_\m =0\quad , \label{eq2}
\ee where $\Phi_{\m\n}$ is symmetric and traceless. These equations
can be obtained from the lagrangian \be {\cal L} =
-\frac{1}{2}(\de_\m\Phi_{\n\r})^2 + (\de\cdot\Phi)_\m^2 -
\frac{m^2}{2}\Phi_{\m\n}^2  \ee only if the constraint
$(\de\cdot\de\cdot\Phi)=0$ is imposed. It is therefore necessary to
include a Lagrange multiplier, that is an auxiliary field $\Phi$,
whose field equation gives $\Phi=0$ and
$(\de\cdot\de\cdot\Phi)=0$\footnote{In the notation of the previous
section, this auxiliary field was the trace of the rank-2 field
itself.}. One therefore considers the lagrangian \be {\cal L}= -
\frac{1}{2}(\de_\m\Phi_{\n\r})^2 - \frac{m^2}{2}\Phi_{\m\n}^2
+(\de\cdot\Phi)_\m^2-\frac{2}{3} [-\frac{1}{2}((\de_\m\Phi)^2+
2m^2\Phi^2) +\Phi_{\m\n}\de^\m\de^\n\Phi ]\quad, \label{lag2} \ee
giving rise to a system of two equations for $\Phi$ and
$(\de\cdot\de\cdot\Phi)=0$, whose determinant \be
\det\left( \begin{array}{cc}2(\frac{1}{2}\Box-m^2) & 1\\
-\frac{1}{2}\Box^2 & (-m^2-\frac{1}{2}\Box) \end{array} \right)=
2m^4 \ee is {\it algebraic} and non-vanishing, and thus admits
$\Phi=0$ and $(\de\cdot\de\cdot\Phi)=0$ as the only solution.

In \cite{singhhagen} it was shown how this can be generalized to any
(integer and half-integer) spin $s$ field. In the case of integer
spin the theory is described in terms of a field $\Phi^{(s)}$,
symmetric and traceless, and a set of auxiliary fields
$\Phi^{(s-k)}$, with $k=2,3,...,s$, again symmetric and traceless.
This is precisely consistent with what we get in the St\"uckelberg
formulation \cite{tesifabio}, where the auxiliary fields are the
components of the lower rank fields that can not be put to zero
fixing the gauge because of the constraint (\ref{epsilon}). It is
interesting to consider in detail the spin 3 case, which is the
first for which the constraint (\ref{epsilon}) is non-trivial. After
gauging away $\phi^{(0)}$, $\phi^{(1)}_\m$ and the traceless part of
$\phi^{(2)}_{\m\n}$, one is left with the trace of $\phi^{(2)}$ and
with the spin 3 field $\phi^{(3)}_{\m\n\r}$. After redefining the
fields, this is the same as a traceless rank 3 field, together with
a rank 1 field (the trace of $\phi^{(3)}$) and a rank 0 field, which
are precisely the auxiliary fields of \cite{singhhagen}. The
procedure of showing that the St\"uckelberg formulation and the one
of ref. \cite{singhhagen} are equivalent was explicitly carried out
for the spin 3 case in ref. \cite{tesifabio}.

Coming back to our equations for arbitrary $s$, if $m =0$ the
situation is the same as the one we encountered for spin 2, since
none of the gauge parameters can be used to gauge away any of the
fields, and therefore all the fields $\phi^{(s-k)}$, $k=0,1,...,s$,
become massless. In order to see this from our equations, one has to
perform recursive field redefinitions, so that eq. (\ref{epsilon})
becomes a traceless condition for the redefined parameters, while
eq. (\ref{fourpointten}) becomes a double-traceless condition for
the redefined fields, whose gauge transformations look exactly like
eqs. (\ref{fourpointeight}) with $m=0$ in terms of the new
parameters.

\subsection{Any $s$ in AdS}

We now want to consider the same system of equations in AdS.
Generalizing the spin 2 case, we are confident that all the fields
can be gauged away up to spin $s-1$, even when
$m=0$\footnote{See~\cite{DESWALD} and~\cite{ZINO} for similar
results.}. We do not worry for the moment about the constraints on
the gauge parameters (\ref{doubletraceps}) and on the fields
(\ref{doubletrace}), we will comment on them at the end of this
subsection.

The idea is the following: we consider the field equation for
$\phi^{(s-1)}$, and we gauge away $\phi^{(s-2)}$ and $\phi^{(s-3)}$
using $\e^{(s-2)}$ and $\e^{(s-3)}$. This implies that only the
fields $\phi^{(s)}$ and $\phi^{(s-1)}$ will appear in the equation,
while all the other fields $\phi^{(s-k)}$, with
$k=4,...,s$ are auxiliary fields. The only gauge invariance left is
the one with respect to the {\it traceless} gauge parameter
$\e^{(s-1)}$, and in AdS it will require the addition of a mass term
for $\phi^{(s-1)}$. We then go to the $m=0$ limit, and we see
whether the mass term we included is the AdS mass or not. If the
mass term is equal to the AdS mass, this means that we have a gauge
symmetry enhancement and our procedure is inconsistent. If the mass
term is different from the AdS mass, this instead means that we can
continue the procedure to $m=0$, and we end up with a massless spin
$s$ and a massive spin $s-1$.

Before we proceed, we first derive the AdS mass for a spin $s$ field
\cite{vasiliev,mikha}. We consider the variation of the equation \bea & &  \Box
\phi^{(s)}_{\m_1 ... \m_s} - s \nabla_{(\m_1} (\nabla \cdot
\phi^{(s)} )_{\m_2 ... \m_s) } + \frac{s(s-1)}{2} \nabla_{(\m_1}
\nabla_{\m_2} \phi^{(s)\l}{}_{\l \m_3 ... \m_s)} \nonumber \\
& & - M_{AdS}^2 \ \phi^{(s)}_{\m_1 ... \m_s} - \tilde{M}_{AdS}^2 \
g_{(\m_1\m_2} \phi^{(s)\l}{}_{\l \m_3 ... \m_s)}=0
\label{masslessspinsads} \eea with respect to \be \d
\phi^{(s)}_{\m_1... \m_s} = s \nabla_{(\m_1} \e^{(s-1)}_{\m_2 ...
\m_s)} \quad , \ee where $\e$ is traceless. The gauge variation of
the derivative part of the equation is \be s [ \Box, \nabla_{(\m_1}
] \e_{\m_2 ...\m_s)} + s (s-1) \nabla_{(\m_1} [ \nabla_{\m_2} ,
\nabla_\r ] \e^\r{}_{\m_3 ...\m_s)} \quad . \ee Substituting the AdS
Riemann tensor, we get \be \frac{s(s-2)(D-1) + s(s-1)(s-4)}{L^2}
\nabla_{(\m_1} \e_{\m_2 ... \m_s)} + \frac{2s(s-1)}{L^2} g_{(\m_1
\m_2} (\nabla \cdot \e)_{\m_3 ... \m_s)} \quad . \ee This
contribution is cancelled by the variation of the AdS mass term in
(\ref{masslessspinsads}) if \be M_{AdS}^2 = \frac{(s-2)(D-1) +
(s-1)(s-4)}{L^2} \quad , \quad \tilde{M}_{AdS}^2 =
\frac{s(s-1)}{L^2} \label{adsmasses}\quad . \ee Since the trace of
the field is zero on shell choosing a suitable gauge \cite{mikha},
the relevant AdS mass term is $M^2_{AdS}$.

We now come back to our problem, and consider the equation for the
spin $s-1$ St\"uckelberg field $\phi^{(s-1)}$ of the previous
subsection, that we denote here by $\chi^{(s-1)}$ in order to adhere
with the notation in sections 1 and 2. The expression that we get by
substituting derivatives with covariant derivatives in (\ref{ddim})
with $k=1$ and by removing all the lower rank fields is \bea & &
\Box \chi^{(s-1)}_{\m_1 ...\m_{s-1}} - (s-1) \nabla_{(\m_1} ( \nabla
\cdot \chi^{(s-1)})_{\m_2 ... \m_{s-1})} \nonumber \\
& & +\frac{(s-1)(s-2)}{2}\nabla_{(\m_1}
\nabla_{\m_2} \chi^{(s-1)\l}{}_{\l\m_3 ... \m_{s-1})} \nonumber \\
& & - m (\nabla \cdot \phi^{(s)} )_{\m_1 ... \m_{s-1}} + (s-1) m
\nabla_{( \m_1} \phi^{(s)\l}{}_{\l\m_2 ...\m_{s-1})}  \quad ,\eea
whose variation with respect to \be \d \phi^{(s)}_{\m_1 ...\m_s} = s
\nabla_{(\m_1} \e_{\m_2 ...\m_s)} \quad , \quad \d
\chi^{(s-1)}_{\m_1 ...\m_{s-1}} = m \e_{\m_1 ...\m_{s-1}} \quad
\label{gaugephichi} \ee is \be m (s-1) [\nabla_{(\m_1}, \nabla_\r ]
\e^\r{}_{\m_2 ... \m_{s-1})}
 = \frac{m (s-1) [ (D-1) + (s -2)]}{L^2} \e_{\m_1 ...\m_{s-1}}
\quad , \ee assuming that $\e$ is traceless. The mass term that
cancels this variation is \be - \frac{(s-1) [ (D-1) + (s -2)]}{L^2}
\chi^{(s-1)}_{\m_1 ...\m_{s-1}} \quad , \label{massterms-1} \ee so
that one ends up with the equation \bea & & \Box \chi^{(s-1)}_{\m_1
...\m_{s-1}} - (s-1) \nabla_{(\m_1} ( \nabla \cdot
\chi^{(s-1)})_{\m_2 ... \m_{s-1})} \nonumber \\
& & +\frac{(s-1)(s-2)}{2}\nabla_{(\m_1}
\nabla_{\m_2} \chi^{(s-1)\l}{}_{\l\m_3 ... \m_{s-1})} \nonumber \\
& & - m (\nabla \cdot \phi^{(s)} )_{\m_1 ... \m_{s-1}} + (s-1) m
\nabla_{( \m_1} \phi^{(s)\l}{}_{\l\m_2 ...\m_{s-1})} \nonumber \\
& &- \frac{(s-1) [ (D-1) + (s -2)]}{L^2} \chi^{(s-1)}_{\m_1
...\m_{s-1}}=0 \quad .\label{chieqmassads}\eea We thus would like to
compare the mass term that we obtained to the AdS mass term, that is
the first of eqs. (\ref{adsmasses}), where $s$ has to be substituted
with $s-1$. They are definitely different, which means that no
symmetry enhancement occurs when $m=0$, and any massive spin $s$
field in the limit of zero mass decomposes into a massless spin $s$
field and a massive spin $s-1$ field. In other words, the new
feature of AdS is the fact that the auxiliary field structure is
preserved for the spin $s-1$ field even when the spin $s$ field
becomes massless. Note that our procedure leaves undetermined a
possible mass term of the form $1/L^2 g_{(\m_1\m_2}
\chi^{(s-1)\l}{}_{\l\m_3 ...\m_s)}$ in eq. (\ref{chieqmassads}),
since $\chi^{(s-1)\l}{}_\l$ is gauge invariant with respect to
(\ref{gaugephichi}). This is not an issue as long as one is focused
on the field equations, since we expect $\chi^{(s-1)\l}{}_\l$ to be
zero on shell using the lower rank equations. Nevertheless, the
whole set of equations should be derivable from a lagrangian once
the correct equations for the auxiliary fields are introduced, in a
similar way to the flat space case.

The difference between the AdS mass term and this mass term (for
simplicity we define $s' =s-1$ from now on) is \be - \frac{2(D-1) +4
(s'-1)}{L^2} \chi^{(s')}_{(\m_1 ...\m_s')} \quad . \ee We therefore
get \be M^2 L^2 = 2(D-1) +4 (s'-1) \quad . \ee In $D=5$ ($d=4$) this
equation becomes \be M^2 L^2 = 4(s'+1) \quad . \ee This is exactly
what we get from the standard relation between mass in AdS and
dimension of the dual operator in the boundary theory
\cite{ferrara,bms}, \be M^2 L^2 = \Delta (\Delta - 4 ) -
\Delta_{min} (\Delta_{min}-4) \quad , \label{identity}\ee with \be
\Delta = s'+ 4 \quad , \qquad \Delta_{min}= s'+2 \quad , \ee which
is exactly the dimension of the corresponding spin $s' $ operator at
vanishing Yang-Mills coupling. For arbitrary dimension $d = D-1$,
$\Delta_{min} = s' + d -2$ represents the unitary bound for the
dimension of the HWS (actually lowest weight state, but we are
physicists) of the UIR with spin $s'$, \ie a totally symmetric rank
$s'$ classically conserved tensor current, and the identity
(\ref{identity}), with 4 substituted with $d$, is satisfied with
$\Delta = s'+d$.

\section{HS multiplets in superspace}

Higher spin (HS) currents $J_{i_1 \dots i_s}$ and their anomalous
terms ${\cal X}_{i_2\dots i_s}$ \be \de^{i_1} J_{i_1 \dots i_s} =
g {\cal X}_{i_2 \dots i_s}\label{anom} \ee occur in ${\cal N}=4$
SYM where, at vanishing coupling $g=0$,  the HS current $J_{i_1
\dots i_s}$ lies in one supermultiplet and the anomalous term
${\cal X}_{i_1 \dots i_{s-1}}$ lies in a different supermultiplet.
In this section we identify examples of such currents and
anomalies and the supermultiplets that contain them. The
supermultiplets containing the HS currents (with spin higher than
2) are singlets under $SU(4)$ and saturate the ${\cal N}=4$
superconformal unitarity bounds in the free theory. In the free
theory, these multiplets assemble together and  with the 1/2 BPS
ultra-short ${\cal N}=4$ supercurrent multiplet form the doubleton
representation of the higher spin group $hs(2,2|4)$
\cite{sezsund,sezsund2,konvaszai,bbms1,bbms2}.

To fix the notation, let us recall that ${\cal N}=4$ SYM consists
of the following elementary fields - all transforming in the
adjoint representation of the gauge group - six scalars
$\phi_{I}\quad I=1\dots 6$ transforming in the vector
representation of the $SO(6)\sim SU(4)$ R-symmetry group, four
complex fermions $\l_{\a A},\bar\l_{\ad}^A\quad A=1\dots 4$
transforming in the spinor representation of $SO(6)$, isomorphic
to the fundamental of $SU(4)$, and the gauge field strength
$F_{ij}$ which is a singlet. Here $\a (\ad)$ are left-handed
(right-handed) Weyl spinor indices and $i,j=1\dots 4$ are vector
indices. We will often represent the field strength tensor as a
bispinor $F_{\a\b}$ defined via the four dimensional
sigma-matrices as \be F_{ij}={1\over 2}
(\s_{ij}^{\a\b}F_{\a\b}+\bar \s_{ij}^{\ad\bd}\bar F_{\ad\bd})\quad
. \ee
 We will also convert between $SO(6)$ indices and $SU(4)$
indices using $SO(6)$  $\Gamma$-matrices, so that
$W_{AB}=\Gamma^I_{AB}W_I$. We use ${\cal N}=4$ (on-shell) Minkowski
superspace to describe the supermultiplets. This superspace has
coordinates $(x^{i},\theta^{\a A},\bar \theta^{\ad}_A)$ and
corresponding supercovariant derivatives $(\de_{i},D_{\a
A},\bar{D}_{\ad}^A)$. These derivatives all (anti)commute with each
other except for the following non-trivial anti-commutator \be
\{D_{\a A},\bar{D}_{\ad}^B\}=i\d_A^B\de_{\a \ad}\quad . \ee

The elementary fields of ${\cal N}=4$ SYM occur in the field
strength superfield $W_I$ whose bottom ($\theta=\bar \theta=0$)
component is the scalar $\phi_I$. We also use the superfields
$\L_{\a A},\bar \L_{\ad}^B$ whose bottom components are the spinors
$\l_{\a A},\bar \l_{\ad}^B$. Supermultiplets can be  formed by
taking particular combinations of gauge invariant products of these
`singleton' superfields acted on by space-time derivatives.

\subsection{Free ${\cal N} =4$}

The superfields we are interested in, which contain HS conserved
currents in the absence of interactions, have the form
\bea
&&{\cal H}^{(s,s)}_{\a_1\dots  \a_{s}\ad_1\dots \ad_{s}}\nonumber\\
&&:=\sum_{k=0}^{s} (-1)^k \left(\ba{c} s\\k \ea \right)^2 Tr(
\de^k_{(\a_1\dots  \a_{k}(\ad_1\dots \ad_{k}}W_I
\de^{s-k}_{\a_{k+1}\dots  \a_{s})\ad_{k+1}\dots \ad_{s})}W_I ) \nonumber\\
&&-  4i\sum_{k=0}^{s} (-1)^k \left(\ba{c} s\\k \ea
\right)\left(\ba{c} s\\k-1 \ea \right) Tr( \de^{k-1}_{(\a_1\dots
\a_{k-1}(\ad_1\dots \ad_{k-1}}\L^A_{\a_k} \de^{s-k}_{\a_{k+1}\dots
\a_{s})\ad_{k}\dots \ad_{s-1}}\bar \L_{\ad_k)
  A}) \label{supercurrents}\\
&&-4\sum_{k=0}^{s} (-1)^k \left(\ba{c} s\\k \ea \right)\left(\ba{c}
s\\k-2 \ea \right) Tr( \de^{k-2}_{(\a_1\dots  \a_{k-2}(\ad_1\dots
\ad_{k-2}}F_{\a_{k-1}\a_{k}} \de^{s-k}_{\a_{k+2}\dots
\a_{s})\ad_{k-1}\dots \ad_{s-2}}\bar F_{\ad_{s-1}\ad_{s})})\nonumber
\eea with $s$ even\footnote{For odd $s$ one gets
conformal descendants \ie total derivatives. Indeed when $s$ is odd
the first line of the above formula for ${\cal
  H}^{(s,s)}$ vanishes.}.
We have converted all Lorentz indices into Weyl spinor indices and
the undotted spinor indices and dotted spinor indices are separately
symmetrised.

The simplest way of obtaining this expression for the supercurrents
is to adapt the expressions found for purely bosonic currents
involving only scalar fields given in~\cite{sezsund,konvaszai,mikha}
\be  {j}^{(s,s)}_{\a_1\dots \a_{s}\ad_1\dots \ad_{s}}
:=\sum_{k=0}^{s} (-1)^k \left(\ba{c} s\\k \ea \right)^2 Tr(
\de^k_{(\a_1\dots \a_{k}(\ad_1\dots \ad_{k}}\phi\
\de^{s-k}_{\a_{k+1}\dots \a_{s})\ad_{k+1}\dots \ad_{s})}\phi ) \ee
using ${\cal N}=4$ analytic superspace~\cite{analytic}. Since this
expression is a primary operator in a conformal field theory in four
dimensions it can be extended to a superconformal primary operator
in ${\cal N}=4$ by simply replacing space-time derivatives
$\de_{\a\ad}$ with analytic superspace derivatives $\de_{\cA \cA'}$
and the scalar $\phi$ with the field strength superfield in analytic
superspace ${\cal W}$ (for notation and a review of superindices in
analytic superspace see~\cite{superindices}) \bea & &  {\cal
  H}^{(s,s)}_{\cA_1\dots  \cA_{s+2}\cA'_1\dots \cA'_{s+2}} \nonumber
  \\
& & :=\sum_{k=0}^{s+2} (-1)^k \left(\ba{c} s+2\\k \ea \right)^2
Tr( \de^k_{(\cA_1\dots  \cA_{k}(\cA'_1\dots \cA'_{k}}{\cal W}\
\de^{s-k+2}_{\cA_{k+1}\dots  \cA_{s+2})\cA'_{k+1}\dots
\cA'_{s+2})}{\cal W} )\ . \label{scurrentsan} \eea Here we have
split the internal  index $A=(a,a')$ so that $a=(1,2)\ a'=(3,4)$
and then the superindices are given as $\cA=(\a,a)$
$\cA'=(\ad,a')$: where $\a,\ad$ are thought of as even indices and
$a,a'$ as odd indices. Symmetrisation of the superindices is
generalised meaning that the $\a$ indices are symmetrised but the
$a$ indices are antisymmetrised. Note that in this expression the
number of analytic superspace derivatives is $s+2$ not $s$.   The
case $s=0$ was first given in an analytic superspace context
in~\cite{AdS/SCFT}. It corresponds to the (in)famous Konishi
multiplet ${\cal K}$ that requires a separate treatment in the
anomaly structure which follows \cite{KONISHI}. In fact the above
expression is also valid for $s=-2$ which corresponds to the
energy-momentum multiplet.

Having found the expression for the supercurrents in analytic
superspace we wish to re-express them in the more familiar Minkowski
superspace. In order to do this we write down the lowest dimension
component of the above operator given by setting two of the unprimed
superindices to be internal and the rest to be external and
similarly for the primed indices. Ie $\cA_i=\a_i, \ i=1\dots s$,
$\cA_{s+1}=a_{s+1},\, \cA_{s+2}=a_{s+2}$ and similarly for the
primed superindices, meaning that we will have $s$ remaining $\a$
and $\ad$ indices as we would expect. Note that we can not have any
more internal indices than two since the internal indices are
anti-symmetrised and there are only two of them. After taking into
account all possible placements of the different types of indices
one obtains~(\ref{supercurrents}).

Other superfields which we are interested in are the short
supermultiplets ${\cal M}_A^{(s-1,s)}$, $\bar{\cal M}^{A(s,s-1)}$
and
  ${\cal N}^{B(s-1,s-1)}_A$
which for $s\geq 2$ have the form\footnote{We thank E. Sokatchev for
  pointing out some missing terms in the definition of ${\cal
    M}_A^{(s-1,s)}$ in the original version of
  this paper. The multiplets ${\cal M}_A^{(s-1,s)}$, $\bar{\cal M}^{A(s,s-1)}$
and
  ${\cal N}^{B(s-1,s-1)}_A$ have since also been given in~\cite{sok} and
  they agree with the expressions given here.}
\bea
{\cal M}_A
%
%
&:=&\sum_{\substack{a,b,c\ge 0\\ a+b+c=s-1}}
A(a,b,c+1)\,Tr\Big(\de^a\bar\L_B[\de^bW_{AC},\de^{c}W^{BC}]\Big)
+\dots\ ,\label{anomaly1}\\[15pt]
{\cal N}^{B}_A\
&:=&\!\!\!\!\sum_{\substack{a,b,c,d\ge 0\\ a+b+c+d=s-1}}
\!\!\!\!B(a,b,c,d)\,
 Tr\Big([\de^aW_{DF},\de^bW^{BF}][\de^cW_{AC},\de^{d}W^{DC}]
\Big) +\dots,
 \label{anomaly2}\eea
where
\bea
A(a,b,c)&:=&
\frac{4\,i\,c^2}{
    \left( a+1 \right)}\frac{\left(  s-1 \right) !\,
    \left( s+1 \right) !}{ \,a!\,b!\,
    c !}\,(k_a+k_b+k_c)\\[10pt]
k_a&:=& \frac{{\left( -1 \right) }^a}
       {a!\,\left( s+1 - a \right) !}\\[15pt]
B(a,b,c,d)&:=&A(a+b,c,d+1){s+1\over s} {a+b\choose a}{b+1\over a+b+2}
%
%
%
%
%
%
\eea
and where we display only the terms involving the combination of fields $\bar \L W^2$ for ${\cal
  M}_A^{(s-1,s)}$ and only terms involving $W^4$ for ${\cal
  N}^{A(s-1,s-1)}_B$. All spinor indices (here suppressed) are
  symmetrised. The dots refer to the remaining
terms involving other combinations of
the fundamental fields and for $\cal N$ they also contain the $SU(4)$
  trace needed to project 
onto the ${\bf 15}$ dimensional irrep. For example ${\cal
  M}_A^{(s-1,s)}$ contains
further terms involving $\bar \L^2 \L,\,W\L\bar F$ and $\bar \L F\bar
  F$. The full expressions will be quite complicated and it would be
  interesting to see if there is a simpler expression for the
  anomalies in analytic  superspace as there was for the supercurrents
  themselves in~(\ref{scurrentsan}).

The superscript $(s,t)$ corresponds to the number $(s)$ of un-dotted
spinor indices and $(t)$ dotted indices which we will largely
suppress. These superfields are subject to the following shortening
conditions in the free theory \bea D^{\a}_A {\cal H}^{}_{\a ...}=0
&\quad &
\bar D^{\ad A} {\cal H}^{}_{\ad ...}=0\label{constr1}\\
D^{\a}_A \bar{\cal M}^{B}_{\a ...} - {1\over 4}\d_A^B D^{\a}_C \bar
{\cal M}^{C}_{...} = 0 &\quad & \bar D^{\ad A} {\cal M}_{B\ad ...}
-{1\over 4}\d^A{}_B \bar D^{\ad C}
{\cal M}_{C\ad ...} =0\label{constr2}\\
D^{\a}_{(A} {\cal M}_{B)\a ...}^{}=0 &\quad&
\bar D^{\ad (A} {\cal \bar M}^{B)}_{\ad ...}=0\label{constr3}\\
D^{\a}_{(C} {\cal N}^{A}_{B)\a ...} -{1\over5}D^{\a}_E {\cal
N}^E_{\a ...(B} \d^A_{C)}=0 &\quad & \bar D^{\ad (C}{\cal
N}^{B)}_{A\ad ...}-{1\over 5}\bar D^{\ad E} {\cal N}_{E\ad
...}^{(B}\d_A^{C)}=0\ .\label{constr4} \eea All of these superfields
lie on the superconformal bounds. It can be explicitly shown that
the supercurrents of eq. (\ref{supercurrents}) satisfy these
constraints. Note that from equation~(\ref{constr1}) it is
straightforward to deduce that the $\theta=0$ component of ${\cal
H}^{(s,s)}$ is a spin $s$ conserved current in the free theory: \be
4i\de^{\a \ad}{\cal H}^{(s,s)}_{\a\ad ...}=\left\{ D^{\a}_{ A}, \bar
D^{\ad A} \right\}{\cal H}^{(s,s)}_{\a\ad ...}=0\quad . \label{cons}
\ee

For $s=0$ the superfield ${\cal H}^{(0,0)} = {\cal K} :=Tr(W_I
W_I)$ satisfies a generalized linearity constraint \cite{AdS/SCFT,
KONISHI} \be {1\over 4} \bar{D}^A_\ad\bar{D}^{B\ad} {\cal K} = 0
\quad , \label{konnonanom} \ee that makes it a semishort multiplet
and in particular implies that the $\theta\bar\theta$ components
(a singlet and a ${\bf 15}$) are conserved vector currents. In
fact all $\theta^s\bar\theta^s$ are conserved spin $s$ currents.

\subsection{Interactions and anomalies}

In the interacting theory the
operators~(\ref{supercurrents},\ref{anomaly1},\ref{anomaly2}) are
given by simply replacing the space-time derivative with the
covariant derivative (using the conventions given in the appendix
of~\cite{drummond}). However the
equations~(\ref{constr1},\ref{constr2}) become anomalous. In fact
the four superfields combine to make a single long superfield.
Equations~(\ref{constr1},\ref{constr2}) become \bea D^{\a}_A{\cal
H}_{\a...}=g {\cal M}_{A ...} &\qquad& \bar D^{\ad A} {\cal M}_{B\ad
...}-{1\over 4}\d_B^A\bar D^{\ad C} {\cal M}_{C\ad ...}= g {\cal
N}^{A}_{B} \\
\bar D^{\ad A} {\cal H}_{\ad ...}=g \bar{\cal M}^{A}_{...} &\qquad &
D^{\a}_A \bar{\cal M}^{B}_{\a ...}-{1\over 4}\d_A^B D^{\a}_C \bar
{\cal M}^{C}_{\a...} =g {\cal N}^{B}_{A} \ . \eea We see that ${\cal
H}$ `swallows up' ${\cal M}, \bar{\cal M} $ and ${\cal N}$ to form a
single superfield which is now unconstrained and hence corresponds
to a long supermultiplet. Indeed we have used these equations to
calculate ${\cal
  M}$ and ${\cal N}$ from ${\cal H}$ in the first place.  Note that equations~(\ref{constr3},\ref{constr4})
still hold: they are now automatically
satisfied. For example~(\ref{constr3}) becomes
\be
D^{\a}_{(A} M_{B)\a...}={1\over g} D^{\a}_{(A} D^{\b}_{B)}{\cal H}_{\a
  \b ...} =0 \quad ,
\ee
which is identically satisfied since ${\cal H}^{(s,s)}$ is totally
symmetric in its
spinor indices (of a given chirality)
and the covariant derivatives anti-commute: $\{D_{\a
  A},D_{\b B}\}=0$.

We are  now in a position to see which multiplets the anomaly sits
in. The conservation condition~(\ref{cons}) becomes
\be
4i\de^{\a \ad} {\cal H}^{(s,s)}_{\a\ad ...}=\left\{ D^{\a}_{ A}, \bar D^{\ad A}
\right\}{\cal H}^{(s,s)}_{\a\ad ...} = g (D^{\a}_{ A}\bar{\cal M}^A_{\a ...}
+ \bar D^{\ad A} {\cal M}_{A\ad ...})
\ee
and so we see that the anomaly ${\cal X}_{i_1...i_{s-1}}$
of~(\ref{anom}) is the real part of
$D^{\a}_{ A}\bar {\cal M}^A_{\a ...}$.

Note that the anomaly ${\cal X}^{(s-1,s-1)}$, though totally symmetric in its
$s-1$ vector indices
(or as many undotted and dotted spinor indices) is not a current itself
even in the absence of interactions. This can be
deduced from conformal symmetry since it does not saturate the
conformal unitarity bounds of the current type.
Indeed ${\cal H}^{(s,s)}$ has spin $s$ and dilation weight $s+2$
and hence saturates the relevant unitary bound. Whereas the anomaly
${\cal X}^{(s-1,s-1)}$ has
spin $s-1$ but dilation weight $s+3= (s-1) + 2 + 2$ and hence does not.

For $s=0$ the ${\cal N}=4$ Konishi anomaly \cite{KONISHI,koniano}
reads \be {1\over 4} \bar{D}^A_\ad\bar{D}^{B\ad} {\cal K} = g
{\cal E}^{AB}\quad , \label{konanom} \ee where in turn  the
superfield \be {\cal E}^{AB} = {1\over 3!} \Gamma_{IJK}^{AB}
Tr(W^I[W^J,W^K]) \ee is 1/8 BPS in the free theory and satisfies
\be {1\over 4} {D}^\a_A D_{B\a} {\cal E}^{CD} = g {\cal
V}^{CD}_{AB}\quad , \ee where finally the superfield \be {\cal
V}^{CD}_{AB} = {1\over 8} \Gamma_{IJ}^{(C}{}_{(A}
\Gamma_{KL}^{D)}{}_{B)}Tr([W^I,W^J][W^K,W^L]) - ...
\ee is a 1/4 BPS multiplet in free theory \cite{quarterbps}. Terms
in dots denote subtraction of $SU(4)$ traces needed to project
${\cal V}_{AB}^{CD}$ onto the ${\bf 84}$ dimensional irrep with
Dynkin labels $[2,0,2]$.

More general shortening conditions that involve the boundary
counterparts of the KK excitations of the HS gauge fields can be
consider as in \cite{osborn,bbms1} but we will not dwell on such
cases.

\section{Conclusions and perspectives}

To conclude we would like to present a synthesis of our results
and attempt to put them in the perspective of {\it La Grande
Bouffe}. In agreement with previous analyses \cite{DESWALD,ZINO},
we have found that, contrary to flat spacetime but consistently
with holography, the relevant St\"uckelberg field for the
spontaneous breaking of a HS gauge symmetry associated to an
originally `massless' spin $s$ field in the AdS bulk is a
genuinely massive spin $s-1$ field. The case $s=1$ is an exception
to the rule while already $s=2$ follows the pattern. The general
strategy was to covariantise the flat space field equations, but
it is also possible to apply the same technique directly to the
lagrangian \cite{DESWALD,ZINO}, whose precise form is fixed by
requiring that it propagates the right degrees of freedom. In flat
space, the procedure was to obtain the $D$-dimensional equations
via dimensional reduction. It would be interesting to see whether
such a derivation is possible for our equations directly in AdS,
along the lines of ref. \cite{siegel}.

We have restricted our attention to the case of totally symmetric
bosonic tensors, although for $AdS_5$ more general possibilities are
allowed. Massive fields of mixed symmetry in $AdS_5$ were discussed
in \cite{metsaevmixsym}, while flat space equations for gauge fields
in arbitrary representations of the Lorentz group were introduced in
\cite{siegzwie} and more recently discussed in \cite{bekboul, mh}.
It would be interesting to extend our results to these cases,
possibly by means of BRS dimensional reduction
\cite{Bekaert:2003uc}. The manifestly supersymmetric analysis of
section 5 shows that at least for the bulk theory dual to ${\cal N}
=4$ it should be possible to relate the Higgs mechanism for mixed
symmetry tensors to the one described in the previous sections.
Moreover, very much as all `conserved' HS current supermultiplets
${\cal H}^{(s,s)}$ become part of a unique $HS(2,2|4)$ doubleton
multiplet, the anomalous terms ${\cal M}^{(s,s-1)}_A$ (and its
conjugate $\bar {\cal M}^{A(s-1,s)}$) and ${\cal N}^{A(s-1,s-1)}_B$
become part of the totally antisymmetric tripleton and `window'
tetrapleton respectively \cite{bbms2}. Free massless field equations
for the doubleton have been derived by Sezgin and Sundell
\cite{sezsund,sezsund2} starting from Vasiliev equations
\cite{vasiliev,hs,brux}. The latter in turn require the introduction
of a master gauge connection $\Omega$, comprising all gauge fields
with $s\ge 2$, and a master scalar field $\Phi$, comprising the low
spin ($s=0$ and $s=1/2$) non-gauge fields as well as the generalized
antisymmetric tensors satisfying a `massive' self-duality
constraint. We thus expect it to be possible to introduce `massive'
master fields for the relevant tripleton and tetrapleton and couple
them to the doubleton in a HS symmetric fashion in AdS very much as
we have done in components. We hope to soon report on this.

Finally it is interesting to observe that the discontinuity of the
massive equations in flat space is reminiscent of the van
Dam-Veltman-Zakharov discontinuity of massive gravity \cite{vdvz}.
In \cite{vdvzads} it was shown that no discontinuity occurs in
AdS, and the fact that massive gravity has a different behaviour
in AdS than in flat space in the massless limit could have some
connection with the results presented in this paper and with the
transition studied in \cite{deconfinement}.

\vskip 1cm

\section*{Acknowledgments}
We would like to thank G. Gabadadze and F. Nitti for discussions
about massive gravity, and C.~Bachas, A. Sagnotti, K.~Skenderis and
E. Sokatchev
for various discussions and suggestions. During completion of this work
M.B. was visiting the String Theory group at Ecole Polytechnique,
E.~Kiritsis and his colleagues are warmly acknowledged for their
kind hospitality.  The work of M.B. was supported in part by
I.N.F.N., by the EC programs HPRN-CT-2000-00122,
HPRN-CT-2000-00131 and HPRN-CT-2000-00148, by the INTAS contract
99-1-590, by the MURST-COFIN contract 2001-025492 and by the NATO
contract PST.CLG.978785. The work of F.R. was supported by a European
Commission Marie Curie Postdoctoral Fellowship, Contract
MEIF-CT-2003-500308.

\end{document}